\documentclass[10pt]{iopart}
\usepackage{bm}
\usepackage{graphicx}
\usepackage{color}
\usepackage{float}
\usepackage{url}

\begin{document}

\title{Low-energy spin dynamics of orthoferrites AFeO$_3$ (A = Y, La, Bi)}

\author{Kisoo Park$^{1,2}$, Hasung Sim$^{1,2}$, Jonathan C. Leiner$^{1,2}$, Yoshiyuki Yoshida$^{3}$, Jaehong Jeong$^{1,2,4}$, Shin-ichiro Yano$^{5}$, Jason Gardner$^{5,6}$, Philippe Bourges$^{4}$, Milan Klicpera$^{7,8}$, Vladim\'{i}r Sechovsk\'{y}$^{7}$, Martin Boehm$^{8}$ and Je-Geun Park$^{1,2}$}
\vspace{10pt}
\address{$^{1}$ Center for Correlated Electron Systems, Institute for Basic Science (IBS), Seoul 08826, Republic of Korea}
\address{$^{2}$ Department of Physics and Astronomy, Seoul National University, Seoul 08826, Republic of Korea}
\address{$^{3}$ National Institute of Advanced Industrial Science and Technologh (AIST), 1-1-1 Umezono, Tsukuba, Ibaraki 305-8568, Japan}
\address{$^{4}$ Laboratoire L\'{e}on Brillouin, CEA, CNRS, Universit\'{e} Paris-Saclay, CEA Saclay, F-91191 Gif-sur-Yvette Cedex, France}
\address{$^{5}$ Neutron Group, National Synchrotron Radiation Research Center, Hsinchu 30077, Taiwan}
\address{$^{6}$ Center for Condensed Matter Sciences, National Taiwan University, Taipei 10617, Taiwan}
\address{$^{7}$ Faculty of Mathematics and Physics, Department of Condensed Matter Physics, Charles University, Ke Karlovu 5, 121 16 Praha 2, Czech Republic}
\address{$^{8}$Institut Laue-Langevin, 71 avenue des Martyrs, CS 20156, 38042 Grenoble Cedex 9, France}
\ead{jgpark10@snu.ac.kr}
\vspace{10pt}
\begin{indented}
\item March 2018
\end{indented}

\begin{abstract}
YFeO$_3$ and LaFeO$_3$ are members of the rare-earth orthoferrites family with \textit{Pbnm} space group. Using inelastic neutron scattering, the low-energy spin excitations have been measured around magnetic Brillouin zone center. Splitting of magnon branches and finite magnon gaps ($\sim$2 meV) are observed for both compounds, where the Dzyaloshinsky-Moriya interactions account for most of this gap with some additional contribution from single-ion anisotropy. We also make comparisons with multiferroic BiFeO$_3$ (\textit{R3c} space group), in which similar behavior was observed. By taking into account all relevant local Dzyaloshinsky-Moriya interactions, our analysis allows for the precise determination of all experimentally observed parameters in the spin-Hamiltonian. We find that different properties of the \textit{Pbnm} and \textit{R3c} space group lead to the stabilization of a spin cycloid structure in the latter case but not in the former, which explains the difference in the levels of complexity of magnon band structures for the respective compounds. 
\end{abstract}

\vspace{2pc}

\submitto{\JPCM}
%\maketitle
\ioptwocol

\section{Introduction}
Magneto-electric (ME) multiferroic materials, in which both magnetic and ferroelectric ordering coexist, have attracted much attention due to the tunable magnetic properties via electric field or vice versa. Such materials also present the possibility of various applications in recording device technology or spintronics \cite{fiebig2005ME, eerenstein2006multiferroic, scott2007memory}. While searching for appropriate candidates is far from trivial, one may consider compounds with weak ferromagnetism (wFM) where the reversal of wFM by 180$^{\circ}$ using electric field has been predicted theoretically \cite{ederer2005WFM}. In many cases, the microscopic mechanism of wFM is either Dzyaloshinsky-Moriya (DM) interaction or single-ion anisotropy (SIA) \cite{dzyaloshinsky1958DM, moriya60anisotropic, bertaut1963magnetism}. In this regard, accurately measuring the values of such quantities in real materials is of considerable importance for future applications.

The rare-earth orthoferrites $R$FeO$_3$ are one of most promising model systems in this regard. The Fe$^{3+}$ ions in all of the $R$FeO$_3$ family undergo an antiferromagnetic transition with T$_N$ ranging from 623 K in $R$=Lu to 738 K in $R$=La. These high transition temperatures are due to a strong nearest-neighbor exchange interaction (J $>$ 4 meV) along the Fe$-$O$-$Fe bond and the large magnetic moment of Fe$^{3+}$ (S = 5/2). Most perovskites of ABO$_3$-type exhibit a cubic $Pm\bar{3}m$ structure at high temperature, and a structural transition occurs upon cooling which lowers the symmetry via tilting of edge-shared BO$_6$ octahedra. $R$FeO$_3$ adopts the \textit{Pbnm} space group at this structural transition, the most frequent structure among the perovskites. Such octahedra tilting to \textit{Pbnm} symmetry can be described by Glazer notation: $a^-a^-c^+$ \cite{glazer1972classification}. Since this structure does not break space inversion symmetry (i.e. \textit{Pbnm} is centrosymmetric), no net polarization in $R$FeO$_3$ is expected. 

\begin{figure}[t]
	\includegraphics[width=\columnwidth,clip]{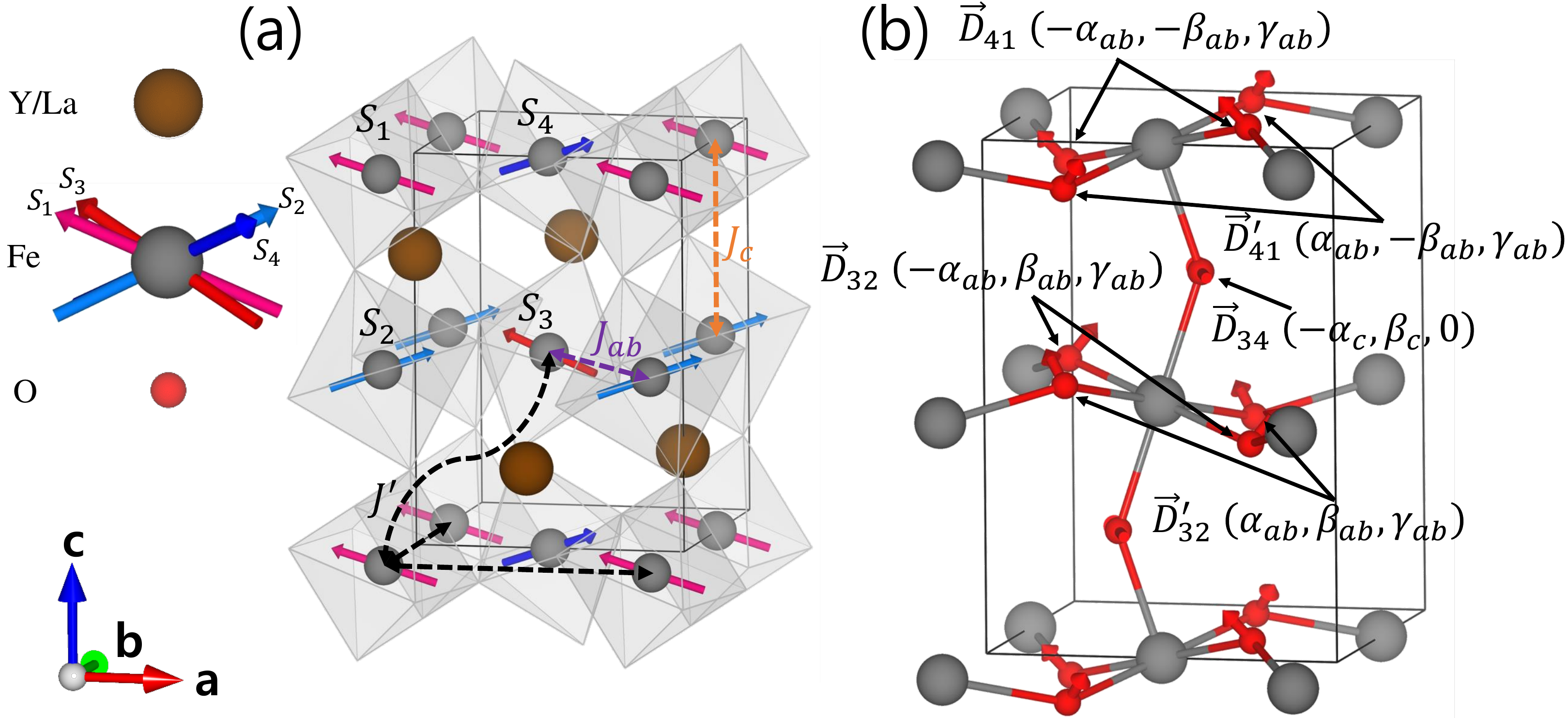}
	\caption{\label{fig1} (Color online) (a) Magnetic structure of YFeO$_3$ and LaFeO$_3$. Exchange interactions bewteen nearest-neighbor bonds along ab-plane ($J_{ab}$) and ${c}$-axis ($J_{c}$), next-nearest neighbor bonds ($J'$) are shown by the dashed arrows, respectively. (b) Local DM vectors (arrows) for nearest neighbors on distorted octahedra.}
\end{figure}

In the case of $R$FeO$_3$, the tilting of FeO$_6$ octahedra is the origin of local DM interaction in this compound (see figure \ref{fig1}). Competition between DM and exchange interactions results in canting magnetic moments \cite{bozorth58origin}. Below T$_N$, all $R$FeO$_3$ adopt a canted antiferromagnetic ground state $\Gamma_4(G_a, A_b, F_c)$ with basic G-type antiferromagnetism along the ${a}$-axis, weak antiferromagnetism along the ${b}$-axis, and weak ferromagnetism along the ${c}$-axis as shown in figure \ref{fig1}(a). Such weak canted magnetic moments were extensively studied both theoretically and experimentally \cite{moskvin2016microscopic, park2008magnetic}. A. S. Moskvin and E. V. Sinitsyn derived a simple formula connecting the canting of magnetic moment and the crystal properties (unit cell parameter, position of oxygen and the bond length), deducing a relation between the $A_y$ and $F_z$ \cite{moskvin1975antisymmetrical}. This theoretical prediction was confirmed for several orthoferrites by the polarized neutron diffraction \cite{plakhtii1981YFO, plakhty83neutron, georgiev1995weak}. For YFeO$_3$, calculated value of $A_y/F_z$ = 1.1 is consistent with the experimental results within errorbars. It is worth noting that in case of $R$FeO$_3$ with magnetic rare-earth ions, there is a magnetic ordering of $R^{3+}$ at low temperature and a spin reorientation transition of Fe$^{3+}$ at intermediate temperatures due to the interaction between $R^{3+}$ and Fe$^{3+}$ ions. Such additional interactions between the two magnetic ions sometimes induces multiferroicity below the spin reorientation transition temperature, and often results in the rotation of Fe$^{3+}$ ions by exchange-striction mechanism \cite{stroppa2010multiferroic, lee2011SmFeO3, zhao2017improper}.

BiFeO$_3$ is the only example that is well-established to exhibit multiferroicity above room temperature. BiFeO$_3$ shares several characteristics with $R$FeO$_3$: it has the similar exchange interaction and the very high antiferromagnetic transition temperature T$_N$ at $\sim$650 K \cite{BFOreview2014}. However, there are also clear contrasts between these two materials such as the distinct rotation of FeO$_6$ octahedra, much of which is due to the lone pair of Bi breaking the inversion symmetry for BiFeO$_3$ unlike the other centrosymmetric $R$FeO$_3$. BiFeO$_3$ has the non-centrosymmetric space group \textit{R3c} coming from the Glazer tilting $a^-a^-a^-$. BiFeO$_3$ exhibits a large polarization with a ferroelectric transition at T$_c$=1100 K \cite{catalan_BFO_2009}. Below T$_N$, an incommensurate spin cycloidal magnetic structure develops along the [1 1 0] direction with an extremely long period of 620 {\AA} and is superimposed on the simple G-type antiferromagnetism \cite{sosnowska95origin}. It was also reported to have a negative magnetostrictive magnetoelectric coupling at T$_N$ \cite{lee2013negative}. Small angle neutron scattering (SANS) experiments revealed a spin density wave (SDW) fluctuation, which is perpendicular to the spin cycloid \cite{ramazanoglu11local}. The local wFM moment made by this fluctuation is cancelled out over the whole cycloid, giving no wFM in bulk BiFeO$_3$. 

The spin-Hamiltonian of BiFeO$_3$ has been extremely well studied both theoretically and experimentally throughout many studies \cite{ohoyama2011high, jeong2012BFO, matsuda12magnetic, jeong2014BFO, fishman15spin}. Recent study on the magnetic excitation spectra over the full Brillouin zone using inelastic neutron scattering (INS) measurements determined the values for the two exchange interactions and the Dzyaloshinskii-Moriya (DM) interaction \cite{jeong2012BFO}. Subsequently, a detailed examination was done on the low-energy region with the observation of the unique island-like feature at 1 meV. Separately, this can also be identified as the peak-and-valley feature in the constant Q-cut graph at the magnetic zone center \cite{matsuda12magnetic, jeong2014BFO}. By employing the full spin Hamiltonian in spin wave calculations, it was further determined that this feature originates from the interplay of the DM interaction and the easy-axis anisotropy \cite{jeong2014BFO}.

The rare-earth orthoferrites have also been previously characterized in the literature, including studies on the spin waves of $R$FeO$_3$ with INS \cite{shapiro_neutron-scattering_1974,koshizuka1980inelastic,gukasov_neutron_1997,mcqueeney2008LFO,hahn14YFO} and Raman spectroscopy \cite{white82light,venugopalan_magnetic_1985,koshizuka1988raman}. Much of the focus in the INS studies was concentrated on the high energy transfer region of the excitation spectra to determine the structural and magnetic interaction strengths. For LaFeO$_3$, only powder INS spectra was reported that confirmed Heisenberg type nearest-neighbor exchange interactions between Fe$^{3+}$ ions \cite{mcqueeney2008LFO}. For YFeO$_3$, a recent INS study successfully measured the overall shape of magnon dispersion up to $\sim$70 meV and deduced the best fit parameters including the nearest- and next nearest-exchange interactions $J_1$ and $J_2$, DM interactions, and SIA \cite{hahn14YFO}. In addition, the low-energy transfer region at the Brillouin zone center was examined by Raman spectroscopy. These Raman measurements for YFeO$_3$ determined the magnon peaks around $\sim$1.4 and 2.2 meV at the $\Gamma$ point \cite{white82light}. Using these data, they also determined the parameters of the spin Hamiltonian of YFeO$_3$. However, the model Hamiltonian used in the above studies needs to be improved as it does not capture all the salient details of \textit{Pbnm} symmetry, in particular the local DM vectors and their relation with the canted ferromagnetic moment. We note that local DM vectors are present even for centrosymmetric space group like \textit{Pbnm} of $R$FeO$_3$, and it is rather poorly understood how this local DM vectors affect the spin waves.

To understand the differences between these two compounds and the role of local DM vectors, it is necessary to quantitatively determine their full spin Hamiltonian. In this work, we have carried out comprehensive studies on the low-energy magnon excitations of YFeO$_3$ and LaFeO$_3$ since this is where effects of DM interaction and SIA are expected to manifest most strongly. We also collected new data of the low-energy spin waves of BiFeO$_3$ focusing on higher momentum resolution. Note that we have purposely selected the nonmagnetic rare-earth YFeO$_3$ and LaFeO$_3$ orthoferrites in order to focus directly on the magnetism of Fe$^{3+}$. Based on the allowed form of the DM interactions in the \textit{Pbnm} symmetry, we have quantified the parameters of the full spin Hamiltonian for YFeO$_3$ and LaFeO$_3$, and reproduced two characteristic features observed in the low-energy magnetic excitation spectra: (1) a finite spin wave gap and (2) splitting of two magnon branches at the zone center. Also, two additional shoulders in constant energy cuts of BiFeO$_3$ have been identified, demonstrating the more complex nature of the magnon branches in comparison to the other orthoferrites.

\section{Experimental Details}
Single crystals of YFeO$_3$ and LaFeO$_3$ with masses of 1.52 and 1.41 g respectively were grown with floating zone furnaces. INS experiments were performed ulitizing the cold-neutron triple axis spectrometer SIKA \cite{wu2016sika} at the Australian Nuclear Science and Technology Organisation (ANSTO). Samples were mounted with their orthorhombic ${b}^*$-axis vertical, such that the wave vectors of the observed spin waves were all confined to the ${a}^*-{c}^*$ plane. Based on the reflection conditions of magnetic Bragg peaks for the \textit{Pbnm} space group, all constant-Q energy scans were carried out along the [H 0 0] direction centered on Q = (1 0 1). The final neutron energy was fixed at 5 meV giving a full width at half maximum (FWHM) energy resolution of 0.106 meV at the elastic position. A beam collimator configuration of $40'-40'-60'-40'$ was used to obtain optimized beam intensity and resolution. A cooled polycrystalline berylium filter was installed to remove the higher-order contamination of the scattered beam. Data were collected at 300 K without a cryostat, and then at 1.5 K with an orange cryostat.  

For BiFeO$_3$, the INS experiments were done with two cold-neutron triple axis spectrometers: 4F2 at Laboratoire Leon Brillouin (LLB) and ThALES at Institute Laue-Langevin (ILL). The data obtained at LLB have already been presented by Jeong et. al. \cite{jeong2014BFO} and reproduced here for comparison and subsequent discussion. In all measurements with 4F2, eight co-aligned single crystals of total mass 1.6 g with 3$^{\circ}$ mosaicity were used. To achieve better momentum resolution, one single crystal with mass of 0.58 g was used in ThALES experiment. Similar with $R$FeO$_3$, BiFeO$_3$ samples were aligned in the ${a}^*-{c}^*$ plane. Using 4F2, energy scans along the [H 0 0] direction centered on Q = (1 0 -1) at T = 16 and 270 K with fixed $k_f = 1.2  \AA^{-1}$. In additional measurements with the ThALES instrument, we have measured the constant-energy (E = 3 meV) cut along the [H 0 0] direction centered on Q = (1 0 -1) at T = 270 K. 

\begin{figure}
	\includegraphics[width=\columnwidth,clip]{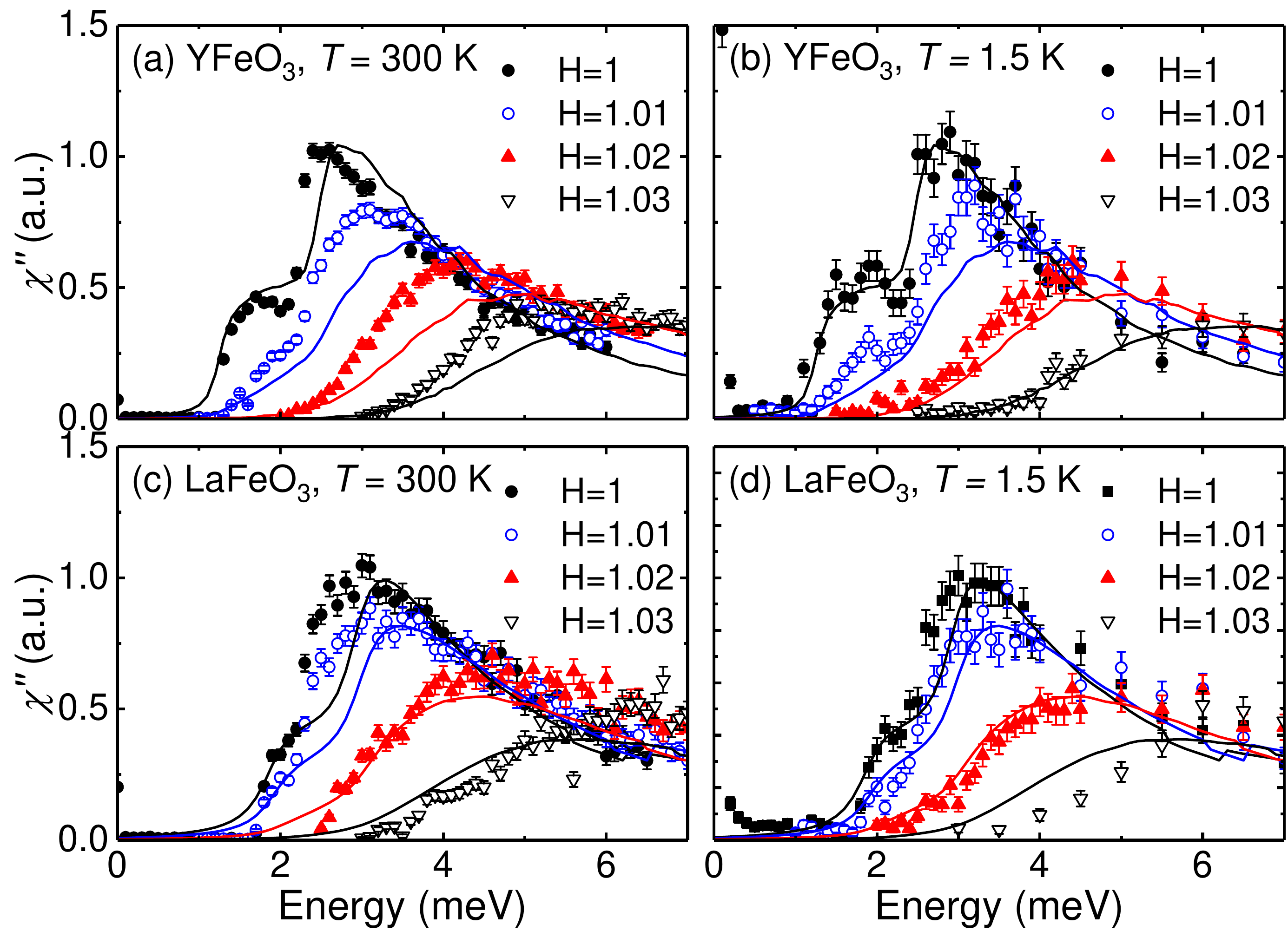}
	\caption{\label{fig2} (Color online) Constant Q-cuts along the [H 0 0] direction centered at Q = (1 0 1) of (a,b) YFeO$_3$ and (c,d) LaFeO$_3$ at T = 300 and 1.5 K. Symbols represent the data points and solid lines denote the convoluted intensity $I(Q,\omega)$ calculated from our simulation as discussed in the text.}
\end{figure}

\begin{figure}
	\includegraphics[width=\columnwidth,clip]{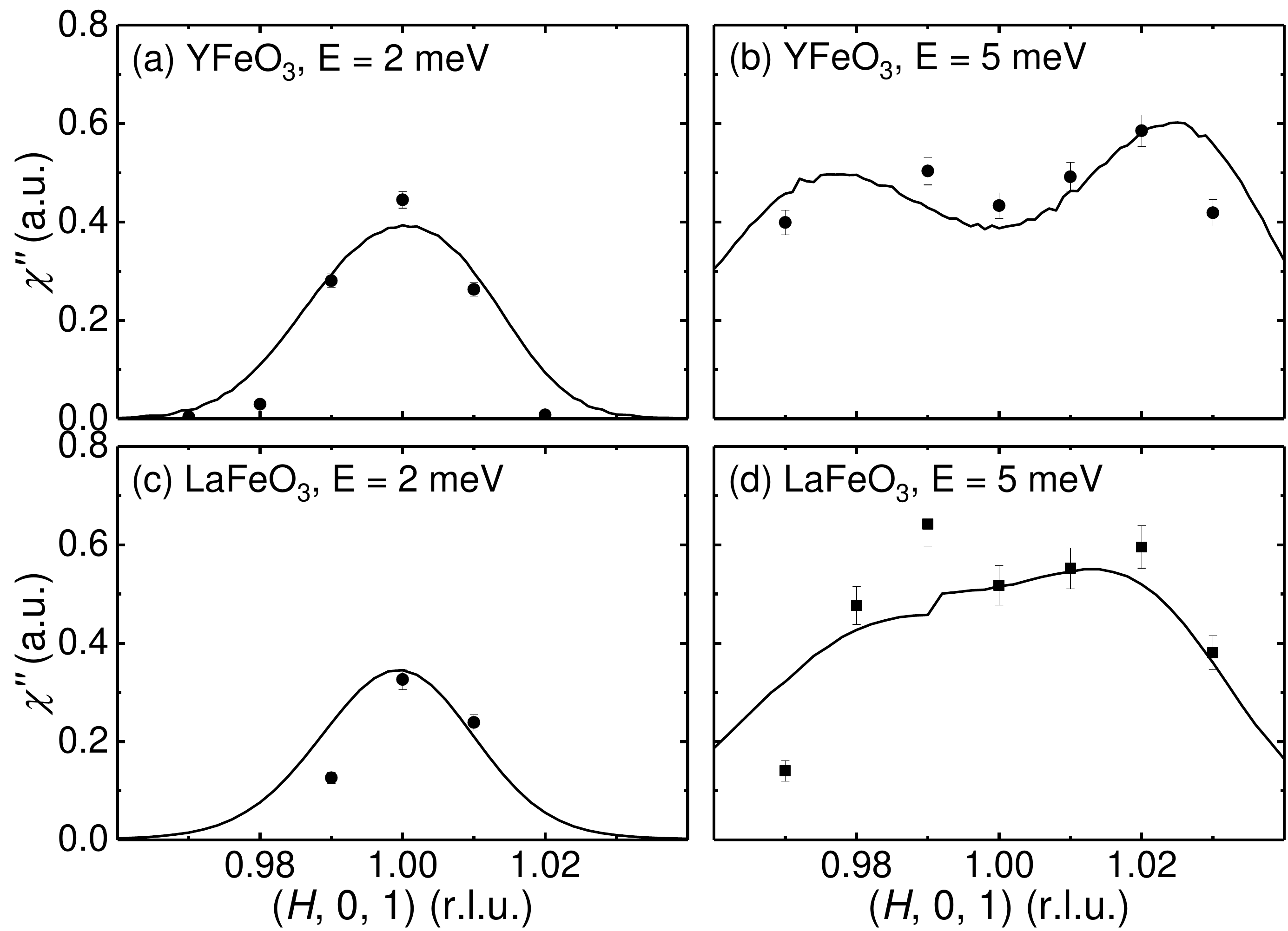}
	\caption{\label{fig3} Constant E-cuts at E = 2 and 5 meV along the [H 0 0] direction centered at Q = (1 0 1) of (a,b) YFeO$_3$ and (c,d) LaFeO$_3$ at T = 300 K. Symbols represent the data points and solid lines denote the convoluted intensity $I(Q,\omega)$ calculated from our simulation as discussed in the text.}
\end{figure}

\begin{figure*}
	\includegraphics[width=\textwidth,clip]{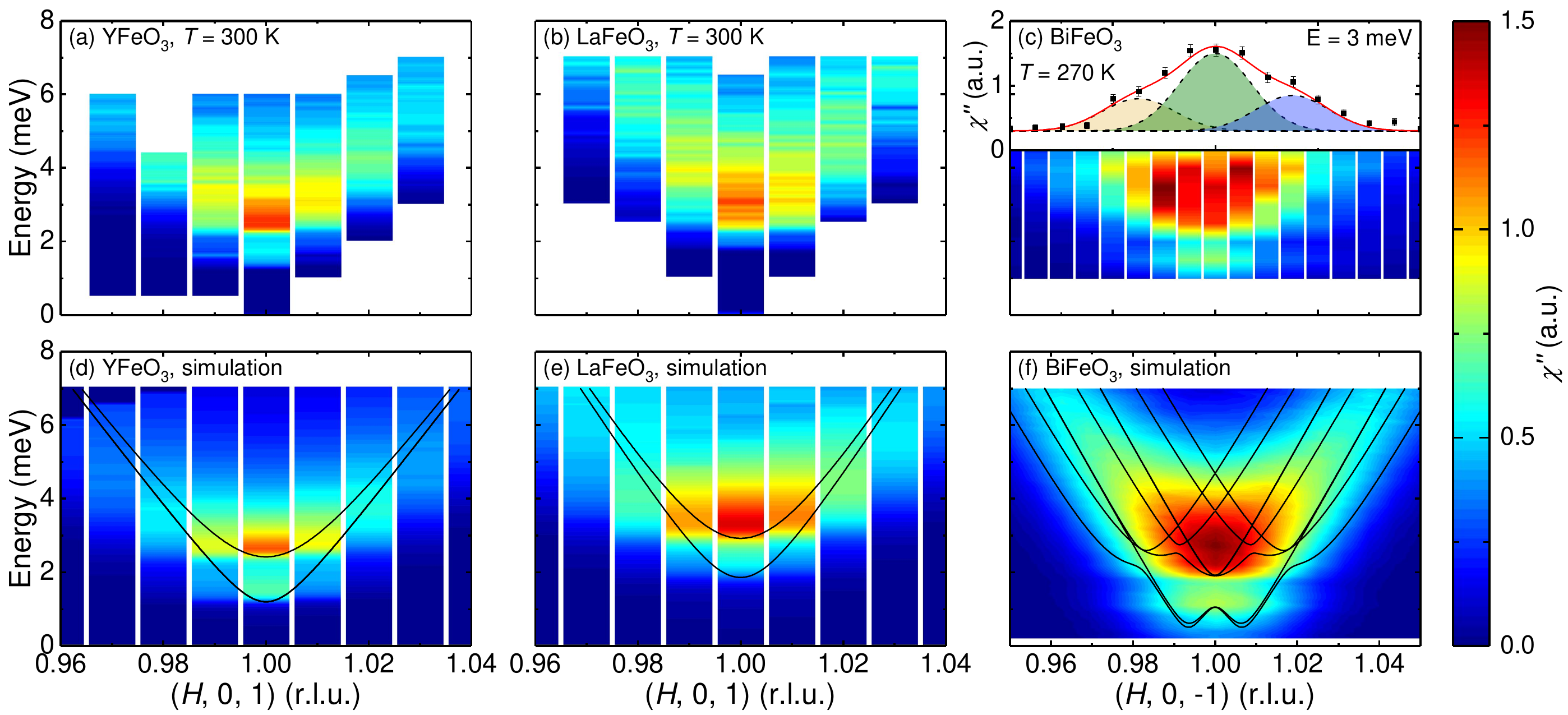}
	\caption{\label{fig4} (Color online) (a-c) Contour plots of the INS intensity of (a) YFeO$_3$ at T = 300 K, (b) LaFeO$_3$ at T = 300 K and (c) BiFeO$_3$ at T = 270 K along the [H 0 0] direction in the reciprocal space. (d-f) theoretically calculated single-magnon dispersion curve and Im[${\chi(Q,\omega)}$] of (d) YFeO$_3$, (e) LaFeO$_3$ and (f) BiFeO$_3$. Experimental data and calculation in (c) and (f) were taken from reference \cite{jeong2014BFO}. The upper part of (c) denotes the constant-energy (E = 3 meV) cut along the [H 0 0] direction centered at Q = (1 0 -1) for BiFeO$_3$ taken from the ThALES spectrometer. Individual Gaussian peaks of single magnon branches are shown as dashed lines with filled area, while the solid line denotes the total sum of all peaks.}
\end{figure*}

\section{Results and Discussion}
Figure \ref{fig2} shows scans in energy transfer at various Q points along the [H 0 0] direction centered at Q = (1 0 1) of YFeO$_3$ and LaFeO$_3$ for T = 300 and 1.5 K. After corrected for the Bose factor, the measured neutron intensities are proportional to the dynamic susceptibility Im[${\chi(Q,\omega)}$]. For both compounds, the defining features in the constant-Q cuts are as follows: (1) a finite spin wave gap of E $\sim$1 meV (YFeO$_3$) and 2 meV (LaFeO$_3$) and (2) two distinct peaks directly above the gap, although the valleys between the two peaks are quite small. The two peaks are, as expected, most distinguishable at Q = (1 0 1), signifying that the magnon branches are split at the magnetic Brillouin zone center. Figure \ref{fig3} denotes the constant-energy transfer graphs of YFeO$_3$ and LaFeO$_3$ for T = 300 K. One can see that the magnetic signals at low energy are separated as two peaks as the energy transfer increases, implying the V-shaped dispersion of the magnetic excitation of YFeO$_3$ and LaFeO$_3$.

In order to fully explain the low-energy magnetic excitations, we employ a minimal spin Hamiltonian of $R$FeO$_3$ to model the experimental data: 

\begin{eqnarray} \label{ham_RE}
{\cal H}& =J_c\sum_{along\,\,c}{\textbf{S}_{i} \cdot \textbf{S}_{j}} +J_{ab}\sum_{ab\,\, plane}{\textbf{S}_{i} \cdot \textbf{S}_{j}} \nonumber \\
        & +J'\sum_{\langle \langle ij \rangle \rangle}{\textbf{S}_{i} \cdot \textbf{S}_{j}} +\sum_{\langle ij \rangle}{\textbf{D}_{ij} \cdot \textbf{S}_{i} \times \textbf{S}_{j}} \\
        & +K_{a}\sum_{i}{\left({S}_{i}^{x} \right)}^{2}+K_{c}\sum_{i}{\left({S}_{i}^{z} \right)}^{2} \nonumber,
\end{eqnarray}

where $J_c$ and $J_{ab}$ represent the nearest-neighbor exchange constants along the ${c}$-axis and the ab plane, respectively. In the previous INS study on YFeO$_3$ \cite{hahn14YFO}, these $J_c$ and $J_{ab}$ were set as same value $J_1$. However, we note that the difference between $J_c$ and $J_{ab}$ can reach up to ~ 10 \% due to Bloch's rule \cite{bloch1966103}, especially in the case of YFeO$_3$. $J'$ denotes the exchange constant along the next-nearest neighbor bonds (see figure \ref{fig1}(a)). The fourth term represents the DM interactions defined on the Fe($i$)-O-Fe($j$) bonds with the antisymmetric relation: ($\textbf{D}_{ij}=-\textbf{D}_{ji}$). Transition ions having a 3d$^5$ configuration such as Fe$^{3+}$ lead to A$_{1g}$ orbital symmetry. Therefore, we may assume that the DM interaction of ferrites can be given by a microscopically derived form ($D_{ij} \propto \hat{x}_i \times \hat{x}_j$) \cite{Keffer1962, moskvin1977exchange}, where $\hat{x}_i$ is the unit vector connecting $i$-th Fe atom and oxygen atom between $i$-th and $j$-th Fe atoms. This means that in the \textit{Pbnm} structure all DM interactions between two adjacent iron atoms may be characterized by five parameters: $\alpha_{ab}$, $\beta_{ab}$, $\gamma_{ab}$, $\alpha_{c}$, $\beta_{c}$ \cite{mochizuki09RMO}, as shown in figure \ref{fig1}(b). The density functional theory (DFT) calculation on LaFeO$_3$ \cite{weingart12noncollinear} shows good agreement with the DM vectors obtained from our structural analysis, supporting this assumption. 

Normalized values of the local DM vectors of YFeO$_3$ and LaFeO$_3$ are shown in table \ref{table1}. We note that the in-plane DM vectors defined in different basal planes, e.g. D$_{41}$ and D$_{32}$, are different along the ${b}$-axis. The result of combining all contributions of adjacent ions is that every Fe$^{3+}$ ion feels a different DM interaction, therefore $global$ DM interactions cannot be as expected defined in this space group. This is an assumption contrary to those used in previous studies on YFeO$_3$ \cite{hahn14YFO, white82light}. 

The last two terms of equation \ref{ham_RE} denote the easy-axis ($K_a$, $K_c$ \textless 0) SIA terms to stabilize the G-type antiferromagnetic order along the $a$-axis and the wFM along the $c$-axis, respectively. With respect to the spin wave theory, SIA is the origin of the spin wave gap at the Brillouin zone center. It is worth noting that the most generalized form of the spin Hamiltonian also includes the symmetric anisotropic exchange interaction, i.e. two-ion anisotropy (TIA). Such TIA terms are formulated as the form $\sum_{ij} \textbf{S}_{i} \Omega_{ij} \textbf{S}_{j}$, where $\Omega_{ij}$ denotes 3 $\times$ 3 symmetric matrix. This is characterized by eight different parameters related to its $Pbnm$ symmetry. This anisotropy, however, seems to be small with the order of D$^2$/J \cite{Shekhtman1992}, and would add unnecessarily too many parameters to our model Hamiltonian. The TIA mostly affects the spin wave gap at the zone center, like the SIA. In that sense, this TIA can be neglected and therefore will not be discussed further in this study.

\begin{table}[!t]
	\centering
	\caption{Normalized components of local DM vectors of YFeO$_3$ and LaFeO$_3$.}
	\begin{tabular*}{0.485\textwidth}{@{\extracolsep{\fill} } cccccc }
		\hline \hline
		& $\alpha_{ab}$ & $\beta_{ab}$ & $\gamma_{ab}$ & $\alpha_{c}$ & $\beta_{c}$ \\ \hline
		YFeO$_3$  & 0.517         & 0.488        & 0.703         & 0.346        & 0.938       \\  
		LaFeO$_3$ & 0.554         & 0.553        & 0.623         & 0.191        & 0.982       \\
		\hline \hline  \label{table1}
	\end{tabular*}
\end{table}

After combining all contributions from the oxygen environments, the four-sublattice magnetic ground state $\Gamma_4(G_a, A_b, F_c)$ of the $R$FeO$_3$ can be stabilized \cite{weingart12noncollinear}. In spherical coordinates, the four spins can be defined using two spin canting angles $\theta$ and $\phi$, which are related to the weak ferro- and antiferro- magnetic moment, respectively.

\begin{equation} \label{spins}
\eqalign{{\textbf{S}}_1& =S\left( -cos\theta cos\phi, -cos\theta sin\phi, sin\theta \right) \cr
{\textbf{S}}_2& =S\left( cos\theta cos\phi, cos\theta sin\phi, sin\theta \right)   \cr
{\textbf{S}}_3& =S\left( -cos\theta cos\phi, cos\theta sin\phi, sin\theta \right)  \cr
{\textbf{S}}_4& =S\left( cos\theta cos\phi, -cos\theta sin\phi, sin\theta \right)}
\end{equation}

Since the spin cantings are very small ($\sim0.5^{\circ}$) for $R$FeO$_3$, we can ignore terms higher than second order with respect to the spin-orbit coupling $\lambda_{SO}$ to obtain the relationship between spin canting angles and the spin Hamiltonian parameters from the ground state energy \cite{plakhtii1981YFO}:
\begin{equation} \label{canting}
\eqalign{\theta = \frac{2\beta_{ab}+\beta_{c}}{4J_{ab}+2J_c+K_c-K_a}, \cr
\phi = -\frac{2\gamma_{ab}}{4J_{ab}-8J'-K_a}}
\end{equation}

\begin{figure}
	\includegraphics[width=0.65\columnwidth,clip]{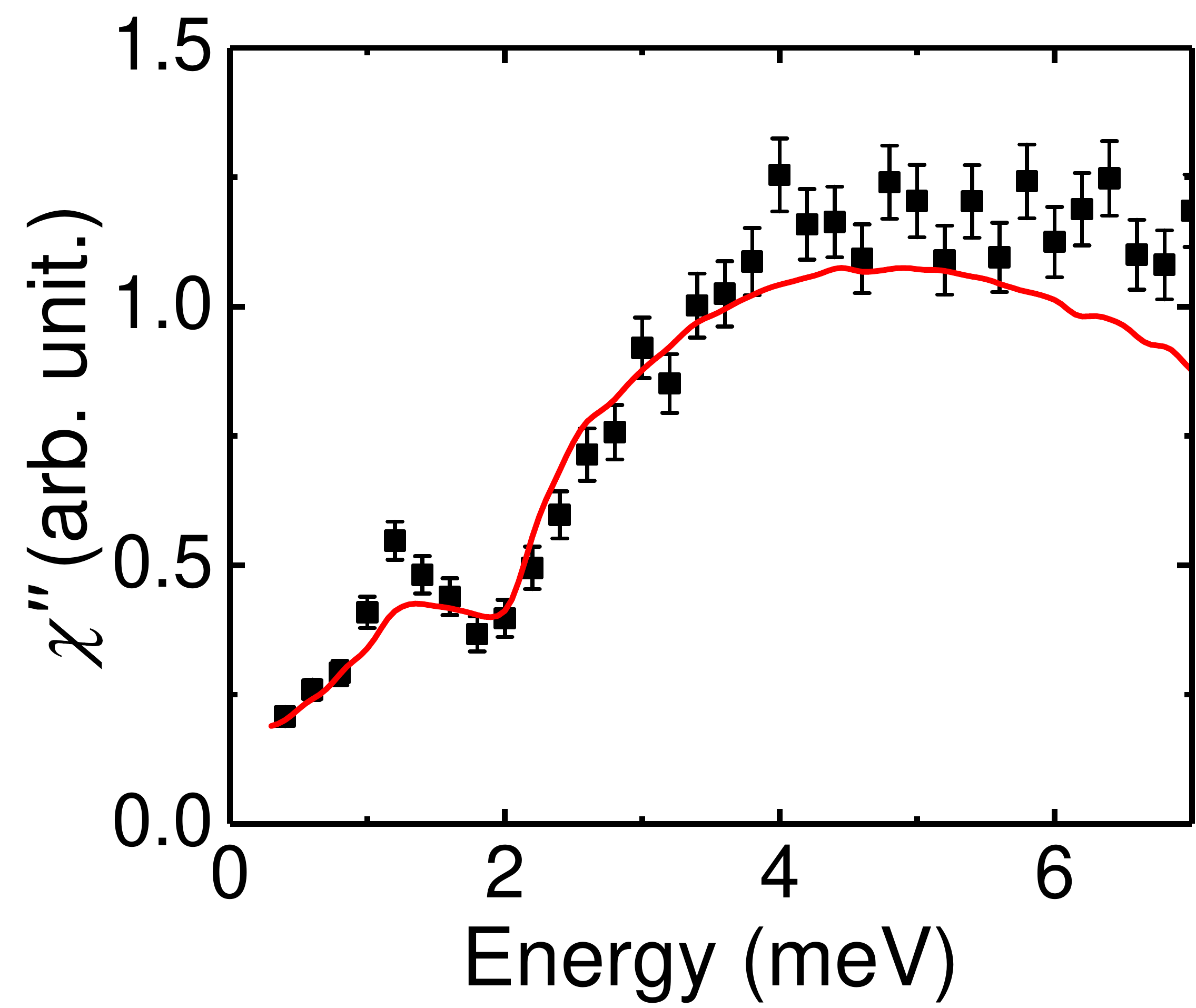}
	\centering
	\caption{\label{fig5} (Color online) Constant Q-cut (black rectangle) at the magnetic Brillouin zone center compared with our simulation (line) of BiFeO$_3$. Experimental data and calculation were taken from reference \cite{jeong2014BFO}.}
\end{figure}

Using this Hamiltonian of $R$FeO$_3$, we tried to find the best fit parameters that reproduce the experimental result well. First, an initial set of parameters was chosen under several constraining conditions. As the Hamiltonian contains many parameters: $J_c$, $J_{ab}$, $J'$, $D_{ab}$, $D_c$, $K_a$ and $K_c$, utilizing all the reasonable initial and constraining conditions is important for determining a reliable set of best fit parameters. Therefore, starting with the previously reported exchange coupling constants $J_1$ = 4.77 and $J_2$ = 0.21 meV derived from high energy INS experiment \cite{hahn14YFO}, $J_c$, $J_{ab}$ and $J'$ were refined. Since only the $J_1$ value of LaFeO$_3$ has been previously reported \cite{mcqueeney2008LFO}, we made the assumption that the $J_1/J_2$ ratio of LaFeO$_3$ is similar with that of YFeO$_3$. This assumption combined with the ratio of T$_N$ of both compounds, yields $J_c$ = $J_{ab}$ = 5.47 and $J'$ = 0.24 meV for LaFeO$_3$. We also used in our analysis the canting angle $\theta$ as derived from polarized neutron diffraction results \cite{plakhtii1981YFO} and magnetization measurements along the $c$-axis \cite{rajendran2006LFOmagnetization}. To obtain the consistency between the spin canting angles and the spin Hamiltonian parameters, equation \ref{canting} was used as one of the constraint conditions.

Secondly, with the chosen initial parameters fitting was performed by a bounded non-linear least squares fit to the experimental data set. Due to the presence of the constraint condition (equation \ref{canting}), $fmincon$ programming solver implemented in MATLAB was used. During the non-linear fit, the theoretical magnon dispersion curve and dynamic structure factor S(Q, $\omega$) have been calculated. We note that the derivation of the analytic form of the dispersion is not easy as the size of Hamiltonian matrix is 8 $\times$ 8. We used SpinW software package \cite{toth2015spinw} to diagonalize the spin Hamiltonian in the Holstein-Primakoff approximation. 

Since the neutron intensity obtained from the triple axis spectrometer is convoluted with the instrumental 4D resolution ellipsoid in the momentum-energy space, the theoretically derived dynamic structure factor should also be convoluted with the resolution ellipsoid for direct comparison with experimental data. The total INS intensity measured by the triple axis spectrometer is given by \cite{zheludev2007reslib}:
\begin{equation} \label{conv}
\eqalign{I(Q_0,\omega_0) &\approx R_0\int d^3Qd\omega S(Q,\omega)\\
&\times exp[-\frac{1}{2}\Delta\wp^{i}M_{ij}(Q_{0}, \omega_{0})\Delta\wp^{j}],}
\end{equation}
where $\textit{Q}_{0}=\textit{k}_{i}-\textit{k}_{f}$ represents the momentum transfer to the sample, $\hbar\omega = E_{i}-E_{f}$ is the energy transfer, $\Delta\wp\equiv(Q-Q_{0},  \hbar(\omega-\omega_{0}))$, and M is a 4 $\times$ 4 matrix defining a 4-dimensional resolution ellipsoid. Based on the geometry of the SIKA beamline and information of the sample, M matrices were calculated via a Cooper-Nathans method in the Reslib library \cite{zheludev2007reslib}. Uniformly sampled 41 $\times$ 41 $\times$ 41 q-points within the ellipsoid were used for a convolution function in the Reslib library. Finally, the convoluted intensity $I(Q,\omega)$ was compared with the experimentally obtained Im[${\chi(Q,\omega)}$] until we get satisfactory convergence of the paramteter. 

\begin{table*}
	\caption{\label{table2}Best fit parameters and spin canting angles used in this work and compared to other work on YFeO$_3$.} 
\begin{indented}
	\lineup
	\centering
	\item[]\begin{tabular}{@{}*{11}{l}}
		\br                              
		& T$_N$ (K) & $J_c$ & $J_{ab}$ & $J'$  & \(|D_{ab}|\) & \(|D_{c}|\) & $K_a$   & $K_c$    & $\theta$ ($^{\circ}$) & $\phi$ ($^{\circ}$)\cr 
		\mr
		YFeO$_3$ (our work)               & 644       & 5.02  & 4.62     & 0.22     & \textbf{0.1206} & \textbf{0.1447}          & \textbf{-0.0091}       & \textbf{-0.0025}   & 0.51   & 0.58\cr
		YFeO$_3$ (reference \cite{hahn14YFO})  & 644       & 4.77  & 4.77     & 0.21     & \00.079 & -      & -0.0055                 & -0.00305 & 0.30        & 0.18\cr 
		LaFeO$_3$ (our work)              & 738       & 5.47  & 5.47     & 0.24     & \textbf{0.130} & \textbf{0.158}          & \textbf{-0.0124}       & \textbf{-0.0037}   & 0.52  & 0.46\cr 
		\br
	\end{tabular}
\end{indented}
\end{table*}

Throughout the above process, the set of parameters that best explain the data was determined. In figure \ref{fig3}, \ref{fig4}(d) and \ref{fig4}(e), the overall V-shapes of the spin-dispersions are modelled accurately by calculations for both compounds. The splitting of magnon branches at the zone center are not as noticeable in the INS data (figure \ref{fig4}(a) and \ref{fig4}(b)). But nevertheless it is fully consistent with theoretical dispersion curves. The constant-Q cuts in figure \ref{fig2}(b)(d) show this consistency more clearly, especially given the tendency for the convoluted $I(Q,\omega)$ to have slightly higher energies due to instrumental resolutions than the calculated energies of the two low-lying magnon branches. For example, the two measured peak positions at Q = (1 0 1) for YFeO$_3$ are at $\sim$1.7 and 2.4 meV, whereas the theoretically calculated magnon energies are at $\sim$1.2 and 2.42 meV. We note that the calculated energies of magnon branches at the magnetic zone center are consistent with Raman data ($\sim$1.4 and 2.2 meV) \cite{white82light}.

The best fit parameters are given in table \ref{table2} together with values for T$_N$ and the spin canting angles. In our work, the values obtained for the DM interactions for YFeO$_3$ are quite different compared to those of Hahn et. al. \cite{hahn14YFO}. We point out two possibilities for this discrepancy: 
\begin{enumerate}
	\item The spin Hamiltonian used in \cite{hahn14YFO} doesn't include DM interaction along the $c$-axis. Since the magnitude of $D_{ab}$ and $D_c$ is similar in $R$FeO$_3$, they should be considered together. 
	\item The canting angles $\theta$ and $\phi$ of YFeO$_3$ used in \cite{hahn14YFO} are much less than the known values ($\sim$0.5$^{\circ}$). Underestimation of the DM vectors is therefore inevitable since they are proportional to the canting angles (equation \ref{canting}). 
\end{enumerate}

The ratio between DM interaction and exchange interaction, $D/J$, is a criterion that indicates the competition between them. A rough estimate for the spin canting angle is given by $\tan^{-1}(D/J)$, and so one can find an approximate value for $D/J$ from equation \ref{canting}. For LaFeO$_3$ the value we obtain for $D/J$ is $\sim$0.026, which is larger than the values obtained from DFT calculations ($\sim$0.018 in reference \cite{weingart12noncollinear}, 0.021 in reference \cite{kim2011DM}). It is also noteworthy that the canting angles of YFeO$_3$ and LaFeO$_3$ are remarkably similar, which is quite unexpected because they have significantly different values for their respective FeO$_6$ octahedra rotation angles. In case of YFeO$_3$, the ratio between canting angles $\theta/\phi \sim$ 1.137 is consistent with previous theoretical and experimental results \cite{moskvin1975antisymmetrical, plakhtii1981YFO, plakhty83neutron, georgiev1995weak}. 

Having said that, the low-energy magnetic excitations of YFeO$_3$ and LaFeO$_3$ have several common features with that of BiFeO$_3$ (see figure \ref{fig4}(c)(f) and figure \ref{fig5}) such as the shoulder-like signal seen below the modes dispersing from the zone center. This feature has been shown to be the result of competition between the three different terms in the Hamiltonian: exchange interaction, DM interaction and SIA. Of course, there is room for this feature to manifest itself in several ways depending on the details. In \textit{Pbnm}, centrosymmetricity and local DM vector constrain $R$FeO$_3$ to have the commensurate 4-sublattice magnetic structure, resulting in the simple V-shape dispersion curves with two of four magnon branches as shown in figure \ref{fig3}. In contrast, all local DM interactions in R3c can be effectively expressed as a global DM interaction along two directions, [1 1 0] and [0 0 1]. Thus, a spin cycloid structure can be stabilized. Furthermore, SDW fluctuations and anharmonicity add more complexity to the structure, making the magnon branches to become more complex. All of these effects combined lead to the distinct behavior of Im[${\chi(Q,\omega)}$] above 4 meV for $R$FeO$_3$ and BiFeO$_3$. 

Since the magnon branches showing up in the INS susceptibility of $R$FeO$_3$ are nearly doubly degenerate: the degeneracy being broken by the DM interaction, there are only two peaks shown in the energy scan. In contrast, as BiFeO$_3$ has many branches of magnons, the scattering intensity remains high above 4 meV at the zone center. INS measurements on BiFeO$_3$ with substantially improved momentum resolution allowed for the observation of the individual magnon branches, as shown in the upper part of figure \ref{fig4}(c). The two peaks at E = 3 meV agree with theoretically calculated magnon dispersion, verifying the 4-fold nature of the magnon dispersions. 

In YFeO$_3$, LaFeO$_3$ and BiFeO$_3$, some care is necessary in choosing the proper relative strengths of the DM interaction and SIA in order to model the spin wave spectra correctly. The SIA in BiFeO$_3$ is not only affected by the DM interaction, but it is also influented by various properties such as ferroelectric distortion and A-site lone-pair effect. This complicated nature of the SIA in BiFeO$_3$ has been explained by new DFT-based calculations (see reference \cite{JHlee_BFO_2015}). The mixing of such parameters yields a temperature dependence of several properties of BiFeO$_3$, e.g. static properties such as the cycloid periodicity and FE distortion as well as the dynamical properties such as the spin wave spectrum. Therefore, the spin Hamiltonian parameters of BiFeO$_3$ are also expected to vary as a function of temperature, which has indeed been observed \cite{jeong2014BFO}.
However, both YFeO$_3$ and LaFeO$_3$ do not show any clear temperature dependence of spin wave spectrum (based on our results collected at T = 300 K and 1.5 K). This implies that the aforementioned static and dynamic properties, and therefore the spin-Hamiltonian parameters of YFeO$_3$ and LaFeO$_3$ remain largely unchanged between 1.5 and 300 K. 

\section{Conclusion}
The low-energy magnon spectra of YFeO$_3$, LaFeO$_3$ and BiFeO$_3$ were studied by our INS experiments. Several features of the magnetic excitation spectra have been explained by the full spin Hamiltonian, which includes the DM interaction and SIA. Best fit parameters of spin Hamiltonian were obtained for YFeO$_3$ and LaFeO$_3$. With the careful quantitative examination of the magnon behavior in these three compounds, we have shown how the relationships between the DM interaction, $J$, and SIA serves as the underlying mechanism driving the spin dynamics. Our study provides a guide for future work on other perovskite systems, in particular with regard to the delicate balance among DM, $J$ and SIA. The values of the magnon mode splitting in most of the other RFeO$_3$ compounds is currently available in the literature \cite{shapiro_neutron-scattering_1974, mcqueeney2008LFO, hahn14YFO, white82light}. Exploiting the relations between these parameters will play a key role in any future implementation of technological applications which utilize Fe$^{3+}$-based perovskites.  

\ack
We thank Joosung Oh, Y. Noda and D. T. Adroja for fruitful discussions. This work was supported by the Institute of Basic Science in Korea (Grant No. IBS-R009-G1). The work of M.K. and V.S. was supported within the program of Large Infrastructures for Research, Experimental Development and Innovation (project No. LM2015050) and project LTT17019 financed by the Ministry of Education, Youth and Sports, Czech Republic.

\section*{References}

\bibliographystyle{iopart-num}
\bibliography{ref}

\end{document}